\documentclass[11pt]{article}
\usepackage[french,english]{babel}
\usepackage[dvips]{graphicx}

\newcommand{\beq}{\begin{equation}}
\newcommand{\eeq}{\end{equation}}
\newcommand{\beqa}{\begin{eqnarray}}
\newcommand{\eeqa}{\end{eqnarray}}

\title{On the finite size corrections to some random matching problems}

\author{G. Parisi\footnote{giorgio.parisi@roma1.infn.it}\ \ and M. Rati\'eville\footnote{matthieu.ratieville@roma1.infn.it} \\{\normalsize Dipartimento di Fisica, INFM and INFN, Universit\`a di Roma 1 {\em La Sapienza}} \\{\normalsize P.le A. Moro, 2 -- 00185 Roma (Italy)}}

\begin{document}

\maketitle

\begin{abstract}
We get back to the computation of the leading finite size corrections to some random link matching problems, first adressed by M\'ezard and Parisi [J. Physique 48 (1987) 1451-1459]. In the so-called bipartite case, their result is in contradiction with subsequent works. We show that they made some mistakes, and correcting them, we get the expected result. In the non bipartite case, we agree with their result but push the analytical treatment further.
\end{abstract}

\section{Introduction}
 
The possibility of investigating some optimization problems using techniques of the field of disordered systems in statistical physics has been recognized for a long time (see \cite{review} for a recent review). Along with the traveling salesman problem and, more recently, K-Sat, one of the problems which got most of the attention is the matching problem, of which two variants have been studied:

\begin{itemize}
\item[(a)] The simple matching problem: consider $2N$ points and a set of 'distances' between them $l_{ij}=l_{ji}$. A matching of these points is a set of $N$ pairs so that each point belong to exactly one pair. The 'length' of such a matching is

\begin{equation}
L_{matching}=\sum_{pair \in matching} l_{pair}.
\end{equation}

One focuses on the properties of the matching of minimal length.
\item[(b)] The bipartite matching problem (or assignment problem), which is as above, except that we split the points into two sub-sets A and B of $N$ points each, and allow only matchings where each pair is made of a point from A and a point from B.
\end{itemize}

Here we are interested in the case where the $l_{ij}$ are independent identically distributed random variables, either uniformly distributed on the interval $[0,1]$ (so-called {\sl flat case}) or distributed with the law $\exp(-l)$ on $[0,+\infty[$ (so-called {\sl exponential case}).

Both the simple and bipartite cases have been investigated in the thermodynamical limit $N\rightarrow +\infty$, where self-averaging of the optimal length occurs. The replica method, in the replica symmetric scheme, yielded predictions for the mean optimal length and the distribution of the lengths of occupied links in the optimal configuration \cite{mezpar2}. This was shown to be equivalent to a cavity approach \cite{mezpar}. Numerical works checked the validity of the results obtained with these techniques \cite{brunetti,houdayer}, and got interested in another quantity, the probability for some given point to be connected to its $k$-th nearest neighbor in the optimal matching. This was also dealt with by an analytical cavity computation \cite{ratieville}. Remarkably, \cite{aldous} confirmed all the above results by rigorous proof.

The stability of the replica symmetric solution was checked in \cite{mezpar3}, yielding as a byproduct the $O(1/N)$ correction to the mean length of the minimal matching. For the assignment problem with flat distances, they found

\beq
\label{faux}
\overline{L}_{min}^{flat}=\frac{\pi^2}{6}-\frac{1}{N}\left(\frac{\pi^2}{12}+2\zeta(3)\right)+o(\frac{1}{N}),
\eeq
where $\overline{\cdots}$ means the average with respect to the distribution of the distances.

This seemed to agree with the numerical simulations at the time \cite{brunetti}. But \cite{conjecture} came up with a conjecture for the assignment problem at any finite $N$: in the exponential case the mean length of the optimal matching would be

\beq
\label{conj}
\overline{L}_{min}^{exp}|_N=\sum_{k=1}^N \frac{1}{k^2},
\eeq
which implies
\beq
\label{vrai}
\overline{L}_{min}^{exp}=\frac{\pi^2}{6}-\frac{1}{N}+o(\frac{1}{N}).
\eeq

The problem is that in the framework of \cite{mezpar3} it is not difficult to prove - as we shall - that

\beq
\label{flatexp}
\overline{L}_{min}^{exp}-\overline{L}_{min}^{flat}=\frac{2\zeta(3)}{N},
\eeq
so that (\ref{vrai}) is not compatible with (\ref{faux}). Beside, other results sustain (\ref{vrai}): more recent numerical simulations \cite{houdayerprivate}, and an allegedly rigorous proof of (\ref{conj}) \cite{dotsenko}.

The main purpose of this article is to show that M\'ezard and Parisi \cite{mezpar3} actually made some mistakes in the computation leading to (\ref{faux}). Correcting them, one does get

\beq
\overline{L}_{min}^{flat}=\frac{\pi^2}{6}-\frac{1}{N} \left(1+2\zeta(3)\right) +o(\frac{1}{N}),
\eeq
which, using (\ref{flatexp}), coincides with (\ref{vrai}).

Apart from some trivial errors due to a confusion about the number of points, $N$ or $2N$, that has already been pointed out \cite{brunetti}, and some typos, there are essentially two mistakes in \cite{mezpar3}:

\begin{itemize}
\item[(i)] they forgot a contribution to the $O(1/N)$ term of $\overline{L}_{min}$, but this is without any consequence as this term turns out to vanish in the zero temperature limit (appendix \ref{derivatives});

\item[(ii)] they made a mistake in the computation of another contribution (appendix \ref{volume}), and this was responsible for the wrong result (\ref{faux}).
\end{itemize}

As the computations carried out in \cite{mezpar3} are quite involved, we have chosen to make this article the most self-contained possible by restating all the necessary steps.

In section \ref{nonbipartite}, we tackle the simple matching problem, which is formally simpler but very similar to the assignment problem. Error (i) is common to both problems, and we deal with it in that section. Moreover we refine the computation of the $O(1/N)$ correction which in \cite{mezpar3} relies on a rough numerical procedure. We give an analytical expression of the correction as the sum of a series. Unfortunately we were not able to sum this series, but it might not be impossible.

In section \ref{bipartite}, we turn to the assignment problem. There we correct error (ii), which is specific to this variant.

\section{Non bipartite case}
\label{nonbipartite}

In this section, for the sake of simplicity, we exclusively consider the {\sl flat case}.

To tackle the problem with the tools of statistical mechanics, one introduces an inverse temperature $\beta$ (to be sent to $+\infty$ in the end) and a partition function

\beq
\label{partition}
Z=\sum_\textrm{all possible matchings} \exp(-N\beta L_{matching}).
\eeq

The scaling factor $N$ ensures a good thermodynamic limit at fixed $\beta$ \cite{vanmez}.

We will not get into the details of the computation of the averaged replicated partition function, because it is quite similar to the bipartite case, for which the derivation can be found in appendix \ref{replication}. Let us just state the result \cite{mezpar3}:
\beqa
\label{replicatedZ}
\overline{Z^n}=\int \prod_\alpha \frac{dQ_\alpha}{\sqrt{2\pi g_{p(\alpha)}/N}} \exp\left(-\frac{N}{2} S[Q] \right)\exp \left(-\frac{1}{4} \sum'_{\alpha, \gamma} \frac{g_\alpha g_\gamma}{g_{\alpha \cup \gamma}^2} Q_{\alpha \cup \gamma}^2 \right),
\eeqa
where $g_p=1/(\beta p)$ and
\beqa
S[Q] &=& \sum_\alpha \frac{Q_\alpha^2}{g_\alpha}-4 \ln z[Q],
\\\nonumber z[Q] &=& \left( \prod_{a=1}^n \int_0^{2\pi} \frac{d\lambda^a}{2\pi} e^{i\lambda_a} \right) \exp \left(\sum_\alpha Q_\alpha e^{-i\sum_{a\in\alpha} \lambda^a} \right).
\eeqa

In the above expressions, $\alpha$ stands for any non empty subset of $\{1,\ldots n\}$ so that the number of $Q_\alpha$ variables is $2^n-1$. For such an $\alpha$, we call $p(\alpha)$ its cardinal number, and use the shorthand notation $g_\alpha$ for $g_{p(\alpha)}$. The notation $\sum_{\alpha,\gamma}'$ means that the summation runs over all the couples $(\alpha,\gamma)$ such that $\alpha \cap \gamma=\emptyset$.

$\overline{L}_{min}$ is nothing but the {\sl intensive} free energy $F/N=-1/(\beta N) \ln Z$ in the limit $\beta \rightarrow +\infty$. It is evaluated by a saddle-point method. The saddle-point equation reads

\beq
\label{sp}
\frac{Q_\alpha}{g_\alpha}=2\frac{\partial \ln z}{\partial Q_\alpha}.
\eeq
It has been solved under the assumption of replica symmetry \cite{mezpar2}: $Q_\alpha^{sp}=Q_{p(\alpha)}^{sp}$. It turns out that the order parameters $Q_p^{sp}$ are not well defined quantities at low temperature, and one can bypass this difficulty by considering the well defined generating function

\beq
\label{G}
G(l)=\sum_{p=1}^{+\infty} \frac{(-1)^{p-1}}{p!}Q_p^{sp} e^{pl},
\eeq
for which (\ref{sp}) translates into
\beq
\label{eqG}
G(l)=-\frac{2}{\beta} \int_{-\infty}^{+\infty} dy K(l+y) e^{-G(y)},
\eeq
where
\beq
\label{defK}
K(u)=\sum_{p=1}^{+\infty} \frac{(-1)^p}{(p!)^2}e^{pu}=-1+\textrm{J}_0(2e^{u/2}).
\eeq

In (\ref{defK}) J$_0$ is the Bessel function of order zero. Note that (\ref{eqG}) can be obtained by direct probabilistic arguments using the cavity method \cite{mezpar}.

The free energy in the thermodynamical limit has been evaluated as \cite{mezpar2}

\beq
\label{thermo}
\frac{F}{N}=\mathop{\lim_{n\rightarrow 0}}_{\beta \rightarrow +\infty} \frac{1}{2\beta n} S[Q_{sp}]=\frac{\pi^2}{12}.
\eeq

The first finite size correction $\Delta F$ is the sum of two terms: the first one, corresponding to the last factor in (\ref{replicatedZ}),
\beq
\label{DF1}
\Delta F^1 = \frac{1}{4Nn\beta} \sum_{\alpha,\gamma}' g_\alpha g_\gamma \left( \frac{Q_{\alpha \cup \gamma}^{sp}}{g_{\alpha \cup \gamma}} \right)^2= -\frac{\zeta(3)}{2N}
\eeq
has been computed in the appendix of \cite{mezpar3}. The second one corresponds to the Gaussian fluctuations around the saddle-point:
\beqa
\nonumber \Delta F^2 &=& \frac{1}{2 \beta N n} \ln \det \left[ \frac{1}{2}\sqrt{g_\alpha}\sqrt{g_\gamma}\frac{\partial^2 S}{\partial Q_\alpha Q_\gamma}\right]
\\\label{DF2} &=& \frac{1}{2 \beta N n} \ln \det T^+_{\alpha\gamma},
\eeqa
where we define the matrices $T^\varepsilon_{\alpha\gamma}$, $\varepsilon=\pm 1$, as follows:
\beqa
\label{defT}
T_{\alpha\gamma}^\varepsilon=\delta_{\alpha \gamma}+\varepsilon\frac{Q_\alpha^{sp} Q_\gamma^{sp}}{2\sqrt{g_\alpha g_\gamma}} -\varepsilon\delta_{\alpha \cap \gamma = \emptyset} \frac{Q_{\alpha \cup \gamma}^{sp}}{g_{\alpha \cup \gamma}}\sqrt{g_\alpha g_\gamma}.
\eeqa

At this stage the introduction of $\varepsilon$ is a useless complication, since for the non bipartite case we only have to consider the case $\varepsilon=+1$. However the case $\varepsilon=-1$ will turn out to be useful in section \ref{bipartite}, where we deal with the bipartite case. 

The computation of  $\Delta F^2$ involves finding the eigenvalues of $(T_{\alpha \gamma}^\varepsilon)$. A vector $(Q_\alpha)$ is eigenvector for the eigenvalue $\lambda$ if  
\beq
\label{eigenproblem}
\forall \alpha,\,\,\,\,\,\,Q_\alpha +\varepsilon \sum_\gamma Q_\gamma \frac{Q_\alpha^{sp} Q_\gamma^{sp}}{2\sqrt{ g_\alpha g_\gamma}} - \varepsilon \sum_{\alpha \cap \gamma = \emptyset} Q_\gamma \frac{Q_{\alpha \cup \gamma}^{sp}}{g_{\alpha \cup \gamma}} \sqrt{g_\alpha g_\gamma}=\lambda Q_\alpha
\eeq

The diagonalization process adopted by \cite{mezpar3} follows the de Almeida-Thouless strategy \cite{at} of considering stable subspaces of increasing dimension. First we look for eigenvectors where no particular replica is distinguished:

\beq
\label{longitudinal}
Q_\alpha=c_{p(\alpha)}.
\eeq

When we plug this into (\ref{eigenproblem}), we see that we have to diagonalize a $n\times n$ matrix $N^{0,\varepsilon}(n)$

\beq
\label{eqforN0}
N^{0,\varepsilon}(n)_{pq}=\delta_{pq}-\varepsilon C_{n-p}^q \frac{Q_{p+q}^{sp}}{g_{p+q}}\sqrt{g_p g_q}+\varepsilon \frac{Q_p^{sp} Q_q^{sp}}{2\sqrt{g_p g_q}} C_n^q \,\,\,\,\,\,\,\,p,q=1,\ldots n,
\eeq
where we use the notation $C_n^p=\frac{n!}{p!(n-p)!}$. This matrix turns into an infinite dimensional matrix $N^{0,\varepsilon}$ when $n\rightarrow 0$:

\beq
\label{N0}
N^{0,\varepsilon}_{pq}=\delta_{pq}-\varepsilon(-1)^q \frac{(p+q-1)!}{(p-1)!q!}\frac{Q_{p+q}^{sp}}{g_{p+q}}\sqrt{g_p g_q} \,\,\,\,\,\,\,\,p,q=1,\ldots +\infty
\eeq
(the last term in (\ref{eqforN0}) does not contribute).

The eigenvalues of $N^{0,\varepsilon}$ have multiplicity 1 in the spectrum of $T^\varepsilon_{\alpha \gamma}$.

Then we look for eigenvectors of $T_{\alpha \gamma}^\varepsilon$ where one replica is distinguished, say $a$:

\begin{displaymath}
\label{onereplica}
Q_\alpha=\left\{ \begin{array}{ll} 
d_{p(\alpha)} & \textrm{if $a\in\alpha$}\\
e_{p(\alpha)} & \textrm{if not} \end{array} \right. .
\end{displaymath}

The orthogonality constraint between this family and the previous one (\ref{longitudinal}) reads $p d_p+(n-p)e_p=0$. So we can choose the only $d_p$ as variables (of which $d_n=0$), and we end up with the diagonalization of an $(n-1)\times (n-1)$ matrix $N^{1,\varepsilon}(n)$

\beqa
\label{eqforN1}
N^{1,\varepsilon}(n)_{pq}=\delta_{pq}-\varepsilon C_{n-p}^q \frac{Q_{p+q}^{sp}}{g_{p+q}}\sqrt{g_p g_q}\frac{q}{q-n} +\varepsilon C_{n-1}^q \frac{Q_p^{sp} Q_q^{sp}} {2\sqrt{g_p g_q}}\frac{q}{q-n} + \varepsilon C_{n-1}^{q-1} \frac{Q_p^{sp} Q_q^{sp}} {2\sqrt{g_p g_q}}\\\nonumber p,q=1,\ldots n-1.
\eeqa

We get $n-1$ eigenvalues $\lambda_1, \ldots \lambda_{n-1}$ (independent on $a$) and the corresponding eigenvectors $u_1^a, \ldots u_{n-1}^a$. The important point is that the $u_i^a,\,\,a=1,\ldots n$ are not linearly independent: $\sum_a u_i^a$ is a vector of the previous family and orthogonal to it, so it has to be 0. Eventually the eigenvalues of $N^{1,\varepsilon}(n)$ have multiplicity $n-1$ in the spectrum of $T_{\alpha \gamma}^\varepsilon$. When $n\rightarrow 0$, $N^{1,\varepsilon}(n)$ becomes an infinite dimensional matrix which happens to be exactly $N^{0,\varepsilon}$.

More generally, one finds the whole spectrum of $T^\varepsilon_{\alpha\gamma}$ by looking for eigenvectors which have $k$ given distinguished replicas:

\begin{displaymath}
Q_\alpha =\left\{ \begin{array}{ll}
0 & \textrm{if $p<k$}\\
d_{p(\alpha)}^i & \textrm{if $\alpha$ contains $k+1-i$ of the distinguished replicas,} \end{array} \right.
\end{displaymath}
where $i$ goes from one to $k+1$.
The orthogonalization with respect to a family where only $k-1$ of these replicas are distinguished reads as a system of equations

\beq
\forall j=0,1,\ldots k-1, \,\,\,\,\,\,\,\,\,\,\, \sum_{r=0}^{k-j} C_{k-j}^r C_{n-k}^{p-(r+j)} d_p^{k+1-(r+j)}=0,
\eeq
whose solution in the $n\rightarrow 0$ limit is
\beqa
\lefteqn{\frac{d_p^1}{p(p+1)\ldots (p+k-1)}=\frac{d_p^2}{(p-k+1)(p+1)\ldots (p+k-1)}}
\\\nonumber & & = \mbox{} \frac{d_p^3}{(p-k+1)(p-k+2)(p+2)\ldots (p+k-1)}=\ldots=\frac{d_p^{k+1}}{(p-k+1)\ldots p}.
\eeqa
(note that this is slightly different from equation (20) in \cite{mezpar3} where there is a typo)

It follows that we can keep the only $d_p^1$ as independent variables, and have to diagonalize a matrix $N^{k,\varepsilon}(n)$, which in the limit $n\rightarrow 0$ is the infinite dimensional matrix

\beq
N^{k,\varepsilon}_{pq}=\delta_{p,q}-\varepsilon (-1)^q \frac{(p+q-1)!(q-1)!}{(p-1)!(q-k)!(q+k-1)!} \frac{Q_{p+q}^{sp}}{g_{p+q}}\sqrt{g_p g_q}\,\,\,\,\,\,\,\,p,q=1,\ldots +\infty.
\eeq

The eigenvalues of $N^{k,\varepsilon}(n)$ have multiplicity $C_n^k-C_n^{k-1}$ in the spectrum of $T^\varepsilon_{\alpha\gamma}$.

The article \cite{mezpar3} prefers using matrices derived from the $N^{k,\varepsilon}$ by some transformations which do not affect the spectrum: after shifting the indices $p$ and $q$ by $k$, then transposing and multiplying each entry by $(-1)^{p+q} \sqrt{\frac{g_{q+k}}{g_{p+k}}} \frac{(q+1)\ldots (q+k-1)}{(p+1)\ldots (p+k-1)}$, one gets the family of matrices

\beq
\label{defM}
M^{k,\varepsilon}_{p,q}=\delta_{p,q}-\varepsilon(-1)^{q+k} \frac{(p+q+2k-1)!}{(p+2k-1)!q!} g_{q+k} \frac{Q^{sp}_{p+q+2k}}{g_{p+q+2k}},
\eeq
where $p,q=0,1, \ldots +\infty$.

We can now proceed to the computation of $\Delta F^2$ (\ref{DF2}):

\beq
\Delta F^2=\Delta F^{2,+}_1+\Delta F^{2,+}_2,
\eeq
where
\beqa
\label{base}
\Delta F^{2,\varepsilon}_1 &=& \lim_{n\rightarrow 0}\frac{1}{2\beta nN} \left[\ln \det N^{0,\varepsilon}(n)+ (n-1) \ln \det N^{1,\varepsilon}(n)\right],
\\\nonumber \Delta F^{2,\varepsilon}_2 &=&  \lim_{n\rightarrow 0} \frac{1}{2\beta nN} \sum_{k\geq 2} (C_n^k-C_n^{k-1}) \ln \det N^{k,\varepsilon}(n).
\eeqa

For $k\geq 2$ one has $(C_n^k-C_n^{k-1}) \sim n(-1)^{k-1} \frac{2k-1}{k(k-1)}$ so that

\beq
\label{F22}
\Delta F^{2,\varepsilon}_2=\frac{1}{2\beta N} \sum_{k=2}^{+\infty} (-1)^{k-1} \frac{2k-1}{k(k-1)} \ln \det M^{k,\varepsilon}.
\eeq

There is a subtlety in the computation of $\Delta F^{2,+}_1$: as the limits of $N^{0,+}(n)$ and $N^{1,+}(n)$ when $n \rightarrow 0$ are the same, one may be tempted to say that in this limit we have a unique family of eigenvalues of multiplicity $n$, and so $\Delta F^{2,+}_1=1/(2\beta N) \ln \det M^{1,+}$. It is what \cite{mezpar3} did, but it is wrong. Actually there is a factor $1/n$ to take into account, so that one also gets the contribution of the derivatives

\beq
\label{der}
\Delta F^{2,+}_1=\frac{1}{2\beta N}\left[ \left\{ \frac{d \ln \det N^{0,+}(n) }{dn}-\frac{d \ln \det N^{1,+}(n)}{dn}\right\}_{n=0}+\ln \det M^{1,+}  \right] .
\eeq

It turns out that the extra term is zero when $\beta \rightarrow +\infty$, but it is not trivial (see appendix \ref{derivatives}). We also show {\em en passant} in this appendix that $\det M^{1,+}$ has a non zero finite limit when $\beta\rightarrow +\infty$ so that we eventually agree with \cite{mezpar3} on the fact that

\beq
\label{DF21}
\Delta F^{2,+}_1=0.
\eeq

As far as the computation of $\Delta F^2_2$ (\ref{F22}) is concerned, the strategy of \cite{mezpar3} consists into translating the infinite dimensional matrices $M^{k,\varepsilon}$ into more tractable integral operators. If $(c_p)$ is an eigenvector of $M^{k,\varepsilon}$, then

\beq
f(x)=\sum_{q=0}^{+\infty} \frac{(-1)^q}{q!} \sqrt{g_{q+k}} c_q e^{(k+q)x-G(x)/2}
\eeq
is an eigenfunction, with the same eigenvalue, of the operator
\beq
M^{k,\varepsilon}(x,y)=\delta(x-y)-\varepsilon (-1)^k A^{k}(x,y),
\eeq
where
\beq
\label{Ak}
A^{k}(x,y)=2 \exp \left(-\frac{G(x)+G(y)}{2}+k(x+y)\right) \sum_{p=0}^{+\infty} \frac{(-1)^p e^{p(x+y)}}{p!(2k+p-1)!}g_{p+k}
\eeq
and reciprocally.

The article \cite{mezpar3}, on the basis of numerical discretization and diagonalization of these operators, argues that the values of $\det M^{k,\varepsilon}(T)$ plotted versus $T\ln k$ fall onto two universal curves, depending on the parity of $k$:

\beq
\label{universal}
\det M^{k,\varepsilon}(T)=\left\{ \begin{array}{ll} 
f_\varepsilon(T\ln k) & \textrm{if $k$ is even}\\
f_{-\varepsilon}(T\ln k) & \textrm{if $k$ is odd} \end{array} \right. .
\eeq

\begin{figure}
\begin{center}
\includegraphics[angle=270,width=10cm]{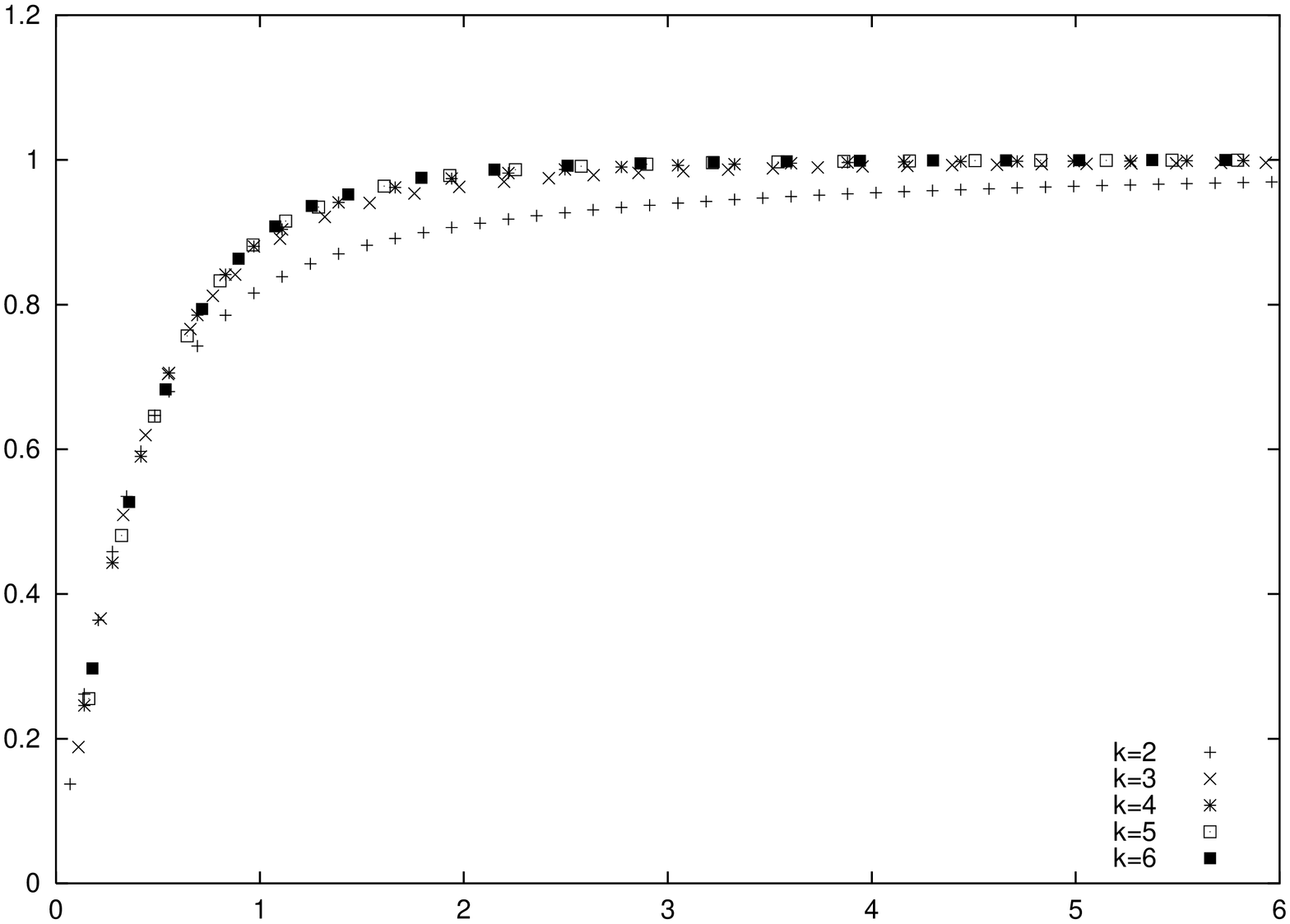}
\includegraphics[angle=270,width=10.1cm]{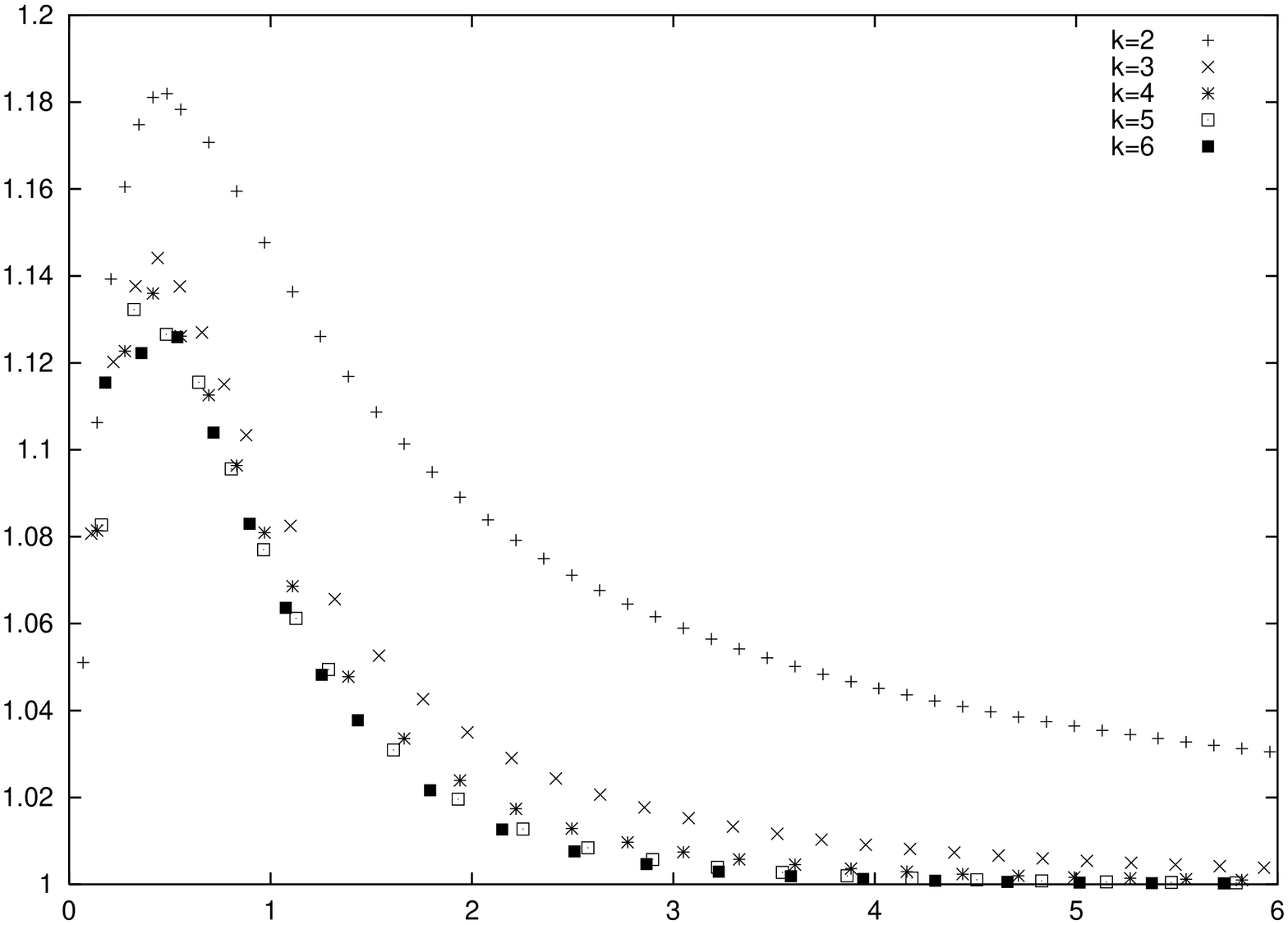}
\caption{The values of $\det M^{k,\varepsilon}(T)$ defined in (\ref{defM}) plotted {\sl versus} $T\ln k$. Above, $\det M^{k,-}$ $k$ odd and $\det M^{k,+}$ $k$ even. Below, $\det M^{k,-}$ $k$ odd and $\det M^{k,+}$ $k$ even}
\label{univ}
\end{center}
\end{figure}

A more accurate numerical analysis showed us that this happens only when $k\rightarrow +\infty$, $T\ln k$ being kept fixed: $f_+$ and $f_-$ are limit functions (see fig. \ref{univ}). Happily this does not change the conclusion that, in the limit $\beta \rightarrow +\infty$,

\beq
\label{corr}
\Delta F^{2,\varepsilon}_2=-\frac{1}{2N} \int_0^{+\infty} dt [\ln f_\varepsilon(t)-\ln f_{-\varepsilon}(t)].
\eeq

In \cite{mezpar3}, this integral is performed by fitting the numerical curves of $f_+$ and $f_-$ by smooth functions, which yields an estimate flawed by a rather rough uncertainty:

\beq
\label{est}
\Delta F^{2,+}_2=\frac{1}{N}(0.47 \pm 0.05).
\eeq

As a consequence, $\Delta F$ itself is known with a bad precision:

\beq
\Delta F=\frac{1}{N}(-0.13 \pm -0.05).
\eeq

It is possible to improve on this. On can explicitly compute the limit of the operator $A^k(x,y)$ when $k\rightarrow \infty$ under the restriction that $t=T\ln k$ remains fixed. In this case one also has $\beta \rightarrow +\infty$. Let us recall that in this limit \cite{mezpar2}

\beq
\label{asymptG}
G(l)=\hat{G}(\beta l)\,\,\textrm{where}\,\,\hat{G}(x)=\ln(1+e^{2x}).
\eeq

So we can write

\beq
A^k(x,y)\sim\frac{2}{\beta\sqrt{(1+\exp(2x/\beta))(1+\exp(2y/\beta))}} f(k,\exp(x+y)),
\eeq
where
\beq
f(k,z)=z^k \sum_{p=0}^{+\infty} \frac{(-1)^p z^p}{p!(2k+p-1)!}\frac{1}{p+k}.
\eeq

The eigenvalues of $A^k(x,y)$ are the same as the ones of the operator

\beq
\frac{\ln k}{t} A^k(\frac{\ln k}{t}u,\frac{\ln k}{t}v)=\frac{2}{\sqrt{(1+\exp(2 u))(1+\exp(2 v))}} f(k,\exp(\frac{\ln k}{t}(u+v))).
\eeq

Let us define

\beq
\label{defg}
g(k,w)=f(k,\exp(w \ln k)).
\eeq

In appendix \ref{limitg} we show that when $k\rightarrow +\infty$, $g(k,w)\rightarrow\Theta(w-2)$ where $\Theta$ is the Heaviside function. So the operator we have to diagonalize is

\beq
\frac{2}{\sqrt{(1+\exp(2 u))(1+\exp(2 v))}} \Theta(\frac{u+v}{t}-2).
\eeq

It is the same as diagonalizing
\beq
H_t(u,v)=\frac{2}{\sqrt{(1+\exp(2 (u+t)))(1+\exp(2(v+t)))}} \Theta(u+v).
\eeq

We have $f_+(t)=\ln \det (I-H_t)$ and $f_-(t)=\ln \det (I+H_t)$. The correction (\ref{corr}) reads
\beqa
\label{serie}
\nonumber \Delta F^{2,+}_2 &=& \frac{1}{2N} \int_0^{+\infty} dt \left[ \ln(\det(I+H_t))-\ln(\det(I-H_t)) \right]
\\ &=& \frac{1}{N} \sum_{p=0}^{+\infty} \frac{I_{2p+1}}{2p+1}
\eeqa 
where 
\beqa
I_p=\int_0^{+\infty} dt\, \textrm{Tr} H_t^p &=& \int_0^{+\infty} dt \int du_1 \ldots du_p H_t(u_1,u_2) \ldots H_t(u_{p-1},u_p) H_t(u_p,u_1)
\\\nonumber &=& 2^p \int_0^{+\infty} dt \int du_1 \ldots du_p \frac{\Theta(u_1+u_2) \ldots \Theta(u_{p-1}+u_p) \Theta(u_p+u_1)}{\left( 1+\exp(2(t+u_1))\right) \ldots \left( 1+\exp(2(t+u_p))\right)}.
\eeqa
(to derive (\ref{serie}) we used the identity $\ln \det=$ Tr $\ln$ and expanded $\ln(\det(I\pm H_t))$ in power series of $H_t$).

Note that the operators $H_t$ have positive and negative eigenvalues. The changes of variables $x_i=\exp(-2u_i)$ and $C=\exp(2t)$ yield

\beq
I_p= \int_0^{+\infty} dt\, Tr H_t^p=\int_1^{+\infty} \frac{dC}{2 C} \int\frac{dx_1}{x_1+C} \ldots \int \frac{dx_p}{x_p+C},
\eeq
where the integration with respect to $x_1,\ldots x_p$ is to be performed over the domain defined by $\forall i, x_i \geq 0$ and $x_i x_{i+1} \leq 1$, $x_p x_1 \leq 1$.

Unfortunately we were not able to compute analytically $I_p$ for a generic $p$. We succeeded in computing exactly the four first terms, and we got an estimate of the fifth one by numerical integration:

\beqa
I_1&=&\frac{\zeta(2)}{4} \sim 0.411234,
\\\nonumber I_2 &=& \frac{\zeta(3)}{2} \sim 0.601028,
\\\nonumber I_3 &=& \frac{3 \zeta(4)}{16} \sim 0.202936,
\\\nonumber I_4 &=& 4 \zeta(5)-\frac{\pi^2 \zeta(3)}{3} \sim 0.193102,
\\\nonumber I_5 &\sim& 0.137098.
\eeqa

A truncated summation of (\ref{serie}) up to the third term gives the following lower bound
\beq
\Delta F^{2,+}_2 > 0.506298/N,
\eeq
which is compatible with the estimate of M\'ezard and Parisi (\ref{est}). Adding $\Delta F_1$ (\ref{DF1}), we get
\beq
\label{prediction}
\Delta F>-0.0947301/N.
\eeq

One can also try a Pad\'e summation to compute $\Delta F_2^{2,+}$. One gets
\beq
N\Delta F \sim I_1+\frac{I_3/3}{1-\frac{I_5/5}{I_3/3}}-\frac{\zeta(3)}{2}=-0.076.
\eeq

\begin{figure}
\begin{center}
\includegraphics[angle=270,width=10cm]{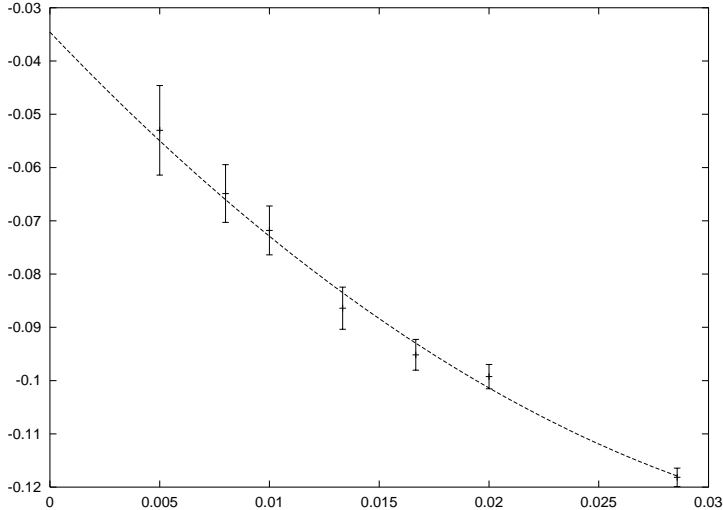}
\caption{$(\overline{L}_{min}-\pi^2/12)N$ {\em versus} $1/N$. The dashed line is the fit (\ref{fit})}
\label{verif}
\end{center}
\end{figure}

To check the validity of (\ref{prediction}), we carried out numerical simulations similar to the ones in \cite{brunetti}, but averaging over more samples and implementing a variance reduction trick \cite{cerf,grassberger}. We used the values $N=35$, 50, 60, 75, 100, 125 and 200, with a decreasing number of samples, from 1200000 downto 300000. The results for $\overline{L}_{min}$ are plotted in fig. \ref{verif}. A quadratic fit

\beq
\label{fit}
\left(\overline{L}_{min}-\frac{\pi^2}{12}\right)N=a+\frac{b}{N}+\frac{c}{N^2}
\eeq
gives $a=-0.0346 \pm 0.0066$, which is compatible with (\ref{prediction}).

\section{Bipartite case}
\label{bipartite}

Now we turn to the assignment problem. To make the comparison with the non bipartite case easier, we prefer using a slightly different convention for the partition function, which amounts to a rescaling of $\beta$: we set

\beq
\label{parti2}
Z=\sum_\textrm{all possible matchings} \exp \left(-\frac{N}{2}\beta L_{matching} \right)
\eeq
instead of (\ref{partition}). However {\sl the reader must bear in mind that in this case $\overline{L}_{min}$ is TWICE the free energy density $F/N=-1/(\beta N) \ln Z$ }.

Moreover, as we want to compare our results with (\ref{vrai}), we must take into account both possible distributions of the distances. In appendix \ref{replication} we sketch the derivation of the averaged replicated partition function both in the case of the flat distribution ($\mu=0$) and in the case of the exponential distribution ($\mu=1$):

\beqa
\label{replicas}
\lefteqn{\overline{Z^n} = \int \prod_\alpha dX_\alpha dY_\alpha \frac{N}{2\pi g_\alpha} \exp \left(-\frac{N}{2} S[X_\alpha,Y_\alpha]\right)}
\\\nonumber & & \times \exp \left(-\frac{1}{2} \sum'_{\alpha, \beta} \frac{g_\alpha g_\beta}{g_{\alpha \cup \beta}^2} (X_{\alpha \cup \beta}^2+Y_{\alpha \cup \beta}^2) \right) \exp \left(-\mu \sum_\alpha X_\alpha^2+Y_\alpha^2 \right)
\eeqa
where
\beq
\label{action}
S[X_\alpha,Y_\alpha]=\sum_\alpha \frac{X_\alpha^2+Y_\alpha^2}{g_\alpha}-2\ln z[X_\alpha-iY_\alpha]-2\ln z[X_\alpha+iY_\alpha].
\eeq

The thermodynamical limit of $F/N$ does not depend on $\mu$, but its correction in $1/N$ does. One can look for a saddle-point of the particular form:

\beq
\label{saddle}
X_\alpha^{sp}=X_{p(\alpha)}^{sp} \,\,\, \textrm{and} \,\,\, Y_\alpha^{sp}=0.
\eeq

$X^{sp}$ satisfy the equation

\beq
X_\alpha=2 g_\alpha\frac{\partial \ln z}{\partial X_\alpha},
\eeq
which is exactly the same as the one for the non bipartite case (\ref{sp}). Hence

\beq
\label{relation}
X_p^{sp}=Q_p^{sp}.
\eeq

The free energy in the thermodynamical limit is the same as in the non bipartite case (\ref{thermo}).

Like in the non bipartite case, the $O(1/N)$ correction to the free energy contains the terms $\Delta F^1$ coming from the last line in (\ref{replicas}), and $\Delta F^2$ coming from the Gaussian fluctuations.

One has $\Delta F^1=\Delta F^1_1 +\mu\Delta F^1_2$, with

\beq
\Delta F^1_1=\frac{1}{2nN\beta}\sum'_{\alpha,\gamma} \frac{g_\alpha g_\gamma}{g_{\alpha\cup\gamma}^2} (Q^{sp}_{\alpha\cup\gamma})^2
\eeq
and
\beq
\label{DF12}
\Delta F^1_2=\frac{1}{nN\beta} \sum_\alpha (Q_\alpha^{sp})^2.
\eeq

We see that $\Delta F_1^1$ is twice the one computed in the case of the simple matching problem (\ref{DF1}). $\Delta F^1_2$, computed in appendix \ref{diff}, happens to be the opposite of $\Delta F^1_1$. So

\beq
\label{resuF1}
\Delta F^1=(\mu-1)\frac{\zeta(3)}{N},
\eeq
which demonstrates our assertion (\ref{flatexp}), confirmed by numerical simulations \cite{houdayerprivate}.

As far as $\Delta F^2$ is concerned, it is easy to generalize the computation of the non bipartite case. We have

\beq
\Delta F^2=\frac{1}{2\beta N n} \ln \left[ \det T_{\alpha\gamma}^+ \det T_{\alpha\gamma}^- \right],
\eeq
where $T_{\alpha\gamma}^\pm$ are the matrices of equation (\ref{defT}). 

Thus we can write $\Delta F^2=\Delta F^{2,+}_1 +\Delta F^{2,+}_2+\Delta F^{2,-}_1 +\Delta F^{2,-}_2$, where the different contributions are defined in (\ref{base}). We have $\Delta F_2^{2,+}+\Delta F_2^{2,-}=0$ (see eq. (\ref{corr})) and we know that $\Delta F_1^{2,+}=0$ (\ref{DF21}).

There is subtlety however in the computation of $\Delta F_1^{2,-}$, as pointed out in \cite{mezpar3}: we run into a problem because $T_{\alpha \beta}^-$ has some zero modes. This is actually no surprise: it arises from the fact that the action (\ref{action}) is left invariant under the transformation
\beq
X_\alpha+i Y_\alpha \rightarrow (X_\alpha+i Y_\alpha) \exp \left(i\sum_{a\in\alpha} \theta_a \right),
\eeq
where $\theta_1, \ldots \theta_n$ are real angles.
The zero modes are the $n$ Goldstone modes of this invariance. As a consequence the saddle-point (\ref{saddle}) is not unique: there is a $n$ dimensional hypersurface of degenerated saddle-points parameterized as follows

\beq
\label{hyp}
X_\alpha^{sp}+iY_\alpha^{sp}=X_{p(\alpha)}^{sp} \exp \left(i\sum_{a\in\alpha} \theta_a \right) \,\,\,\,\,\,\,\,0\leq\theta_i\leq 2\pi.
\eeq
The kernel of $T_{\alpha\beta}^-$ is spanned by the $n$ vectors $\xi_i$ of components

\beq
\xi_i^\alpha=\frac{\partial Y_\alpha^{sp}}{\partial \theta_i}=\left\{ \begin{array}{ll}
X_{p(\alpha)}^{sp} & \textrm{if $i\in\alpha$}\\
0 & \textrm{if not} \end{array} \right. .
\eeq
which have the replica $i$ distinguished. So $N^{k,-}(n)$ for $k\geq 2$ has no zero mode. Only $N^{0,-}(n)$ and $N^{1,-}(n)$ have a zero eigenvalue each.
Thus $\Delta F_1^{2,-}$ is to be computed leaving aside the zero modes, and one has to take into account a new contribution $\Delta F^3$ to the free energy corresponding to the volume of the orbit (\ref{hyp}).

As far a the computation of $\Delta F_1^{2,-}$ is concerned, we refer to appendix (\ref{derivatives}), where we showed that $\Delta F_1^{2,+}=0$. It is easy to see that the proof is exactly the same for $\Delta F_1^{2,-}$: there are only some sign reversals (in particular one has $1+\lambda_k$ instead of $1-\lambda_k$ in the denominator of (\ref{perturb})), and the exclusion of the zero modes (corresponding to $k=1$ in (\ref{polynomial})) has no consequence because the key property (\ref{zero}) holds for each eigenvalue of $I(x,y)$. So $\Delta F_1^{2,-}=0$, and
\beq
\label{resuF2}
\Delta F^2=0.
\eeq

Let us now turn to the computation of the volume of the hypersurface defined by (\ref{hyp}). It is where M\'ezard and Parisi made a mistake: they computed this quantity without taking into account the fact that they carry out the diagonalization in another system of coordinates. To make things clearer let us rewrite (\ref{replicas}) as

\beqa
\lefteqn{\overline{Z^n} = \int \prod dU_\alpha dV_\alpha \frac{N}{2\pi} \exp \left( -\frac{N}{2} S[\sqrt{g_\alpha} U_\alpha,\sqrt{g_\alpha} V_\alpha]\right)}
\\\nonumber & & \times \exp \left(-\frac{1}{2} \sum'_{\alpha, \gamma} \frac{g_\alpha g_\gamma}{g_{\alpha \cup \gamma}} (U_{\alpha \cup \gamma}^2+V_{\alpha \cup \gamma}^2) \right) \exp \left(-\mu \sum_\alpha g_\alpha (U_\alpha^2+V_\alpha^2) \right).
\eeqa

It is in the variables $(U_\alpha,V_\alpha)$ that M\'ezard and Parisi have chosen to diagonalize: indeed the matrix $T^{\pm}_{\alpha\gamma}$ is half the Hessian matrix of $S[\sqrt{g_\alpha} U_\alpha,\sqrt{g_\alpha} V_\alpha]$. So the volume of the hypersurface is to be computed in these same variables, not in $(X_\alpha,Y_\alpha)$ as they did. The saddle-points coordinates are related by

\begin{eqnarray}
\label{hyp2}
\left\{ \begin{array}{ll}
U_\alpha^{sp}&=X_\alpha^{sp}/\sqrt{g_p}
\\ V_\alpha^{sp}&=Y_\alpha^{sp}/\sqrt{g_p} \end{array} \right. . 
\end{eqnarray}

The computation of the correct volume is done in appendix \ref{volume}. We find

\beq
\label{resuF3}
\Delta F^3=-\frac{1}{2N}.
\eeq

Collecting the pieces (\ref{resuF1}),(\ref{resuF2}),(\ref{resuF3}), we get the expected result (\ref{vrai}).

\section{Conclusion}

By fixing the mistakes made by \cite{mezpar3} in the computation of the $O(1/N)$ correction to the mean minimum length in the random assignment problem, we removed any inconsistency among the corpus of results about this problem. This gives further evidence that the replica approach, in its simplest symmetric ansatz, exactly solves the problem, and remains a valuable tool to gain insight on such quantities as this finite size correction which, despite recent dramatic progresses in the rigorous approach \cite{aldous}, still resist a mathematical treatment.

\bigskip

\begin{center}
{\bf ACKNOWLEDGMENTS}
\end{center}

We acknowledge very useful discussions with A. Cavagna, J. Houdayer and M. M\'ezard.

\bigskip

\begin{center}
{\bf APPENDIX}
\end{center}

\appendix

\section{Non bipartite case}

\subsection{Contribution of the derivatives in $\Delta F^2_1$ (\ref{der})}
\label{derivatives}

Our purpose here is to show that

\beq
\label{goal}
\Gamma=\lim_{\beta \rightarrow +\infty} \frac{1}{2\beta N}\left[ \frac{d \ln \det N^{0,+}(n) }{dn}-\frac{d \ln \det N^{1,+}(n)}{dn}\right]_{n=0}
\eeq
equals 0.

We start from the equations (\ref{eqforN0}) and (\ref{eqforN1}). Writing $\frac{q}{q-n}=1+\frac{n}{q}+o(n)$ it is easy to see that $N^{1,+}(n)-N^{0,+}(n)=n\Delta^1+n\Delta^2+o(n)$, where $\Delta_1$ and $\Delta_2$ are the following infinite dimensional matrices

\beqa
\Delta_{pq}^1 &=& \frac{(-1)^{q+1}}{q} \frac{(p+q-1)!}{(p-1)!q!}\frac{Q_{p+q}}{g_{p+q}}\sqrt{g_p g_q}
\\\Delta_{pq}^2 &=& \frac{(-1)^q}{q} \frac{Q_p^{sp} Q_q^{sp}} {2\sqrt{g_p g_q}}\,\,\,\,\,\,\,\,\,\,\,\,\,\,p,q=1,\ldots +\infty.
\eeqa

It is more convenient to use an integral operator formalism. We set

\beq
\label{eq1}
I(x,y)=\frac{2}{\beta} \exp \left( -\frac{G(x)+G(y)}{2} \right) K(x+y),
\eeq
where $K$ is the function defined in (\ref{defK}).

If $(c_p)$ is an eigenvector of $N^{0,+}_{pq}$ (\ref{N0}) for the eigenvalue $1-\lambda$ then $f$ is an eigenfunction of $I(x,y)$ for the eigenvalue $\lambda$, where

\beq
f(x)=\sum_{q=1}^{+\infty} \frac{(-1)^q}{q!} \sqrt{g_q} c_q e^{qx-G(x)/2}.
\eeq

One can easily check that the operators corresponding to the matrices $\Delta^1$ and $\Delta^2$ are respectively

\beqa
\label{eq2}
\Delta^1(x,y) &=& -\frac{2}{\beta}e^{-G(x)/2}e^{G(y)/2} \int_y^{+\infty} K(t+x) e^{-G(t)} dt,
\\ \Delta^2(x,y) &=& -G(x)e^{-G(y)/2}e^{-G(x)/2}.
\eeqa

Thus (\ref{goal}) can be evaluated by the standard result of first order perturbation theory

\beq
\label{perturb}
\Gamma=-\frac{1}{2N}\sum_k \frac{\langle k|\Delta^1|k \rangle +\langle k|\Delta^2|k \rangle}{\beta (1-\lambda_k)},
\eeq
where the $|k \rangle$ are the normalized eigenvectors of $I(x,y)$ (\ref{eq1}), and the $\lambda_k$ are the corresponding eigenvalues.

Below we show that
\beq
\label{zero}
\frac{\langle k|\Delta^1|k \rangle+\langle k|\Delta^2|k \rangle}{\beta}=0
\eeq
for each $k$ when $\beta \rightarrow +\infty$, and that the $\lambda_k$ have finite limits, different from 1, so that $\Gamma=0$.

Let us consider an eigenfunction $f$ of $I(x,y)$:

\beq
\label{eigen}
\int dy \, I(x,y) f(y)=\lambda f(x).
\eeq

We make the substitution

\beq
f(x) e^{G(x)/2}=P(\frac{1}{1+\exp(2x/\beta)}).
\eeq

In the $\beta \rightarrow +\infty$ limit we can use (\ref{asymptG}) and after some changes of variables we see that (\ref{eigen}) can be restated as
\beqa
\nonumber \forall v \in [0,1], \,\,\,\,\,\, \int_{0}^1 du \frac{P(u)}{1-u} K\left[\frac{\beta}{2}\ln\left(\frac{(1-u)(1-v)}{uv}\right)\right]=\lambda P(v)
\\\label{indepb} \textrm{i.e.}\,\,\,\,\,  -\int_{1-v}^1 du \frac{P(u)}{1-u}=\lambda P(v),
\eeqa
because $K(\beta z)=-1$ if $z\geq 0$, 0 otherwise. Note that $P(0)=0$.

The eigenproblem (\ref{indepb})  is $\beta$ independent. We found that its eigenvalues are $\lambda_k=(-1)^k/k$, $k=1,2,\ldots +\infty$. The corresponding eigenfunctions are polynomial of degree $k$:
\beq
\label{polynomial}
P_k(u)=\sum_{p=0}^k a_{p,k} u^p \,\,\,\, \textrm{with} \,\,\,\, a_{p,k}=(-1)^p\frac{k^2 (k^2-1) \ldots (k^2-(p-1)^2)}{(p!)^2}.
\eeq

For the computation of $\langle k|\Delta^1|k\rangle$, it is simpler not to use this explicit form. By derivation of (\ref{indepb}), we get

\beq
\label{eqP}
P(1-v)=\lambda v P'(v).
\eeq

Combining (\ref{eq1}), (\ref{eq2}) and (\ref{eigen}), we have

\beq
\int \Delta^1(x,y) f(x)dx = -\lambda e^{G(y)/2} \int_y^{+\infty} f(t) e^{-G(t)/2},
\eeq
so that
\beqa
\int\int \Delta^1(x,y) f(x)f(y)dx dy &=&\lambda \frac{\beta^2}{4} \int_0^1 du \frac{P(u)}{u(1-u)} \int_u^1dv \frac{P(v)}{1-v}
\\\nonumber &=&-\lambda^2 \frac{\beta^2}{4}\int_0^1 du \frac{P(u)}{u(1-u)}P(1-u) \,\,\,\,\,\,\, \textrm{by (\ref{eqP})}.
\eeqa

This can be further simplified:

\beqa
\int_0^1 du \frac{P(u)P(1-u)}{u(1-u)}&=&\int_0^1 du P(u)P(1-u)\left(\frac{1}{u}+\frac{1}{1-u} \right)
\\\nonumber &=&2\int_0^1 du \frac{P(u)P(1-u)}{u}
\\\nonumber &=&2\lambda \int_0^1 du \, P(u) P'(u)
\\\nonumber &=&\lambda P(1)^2,
\eeqa
so that
\beq
\int\int \Delta^1(x,y) f(x)f(y)dx dy=-\lambda^3 \frac{\beta^2}{4} P(1)^2.
\eeq

The function $f$ is {\sl a priori} not normalized so that the above quantity is to be divided by

\beq
\int f(x)^2 dx = \frac{\beta}{2} \int_0^1 du \frac{P(u)^2}{1-u}.
\eeq

We eventually get

\beq
\langle k|\Delta^1|k \rangle=-\lambda^3\frac{\beta}{2} P(1)^2 \left[\int_0^1 du \frac{P(u)^2}{1-u}\right]^{-1}.
\eeq

Now we compute $\langle k|\Delta^2|k \rangle$: thanks to (\ref{eqG}),
\beqa
\Delta^2(x,y)&=&\frac{2}{\beta}\int dt K(t+x) e^{-G(t)} e^{-G(x)/2} e^{-G(y)/2}
\\\nonumber &=&\int dt I(x,t) e^{-G(y)/2} e^{-G(t)/2},
\eeqa
hence
\beq
\int \Delta^2(x,y) f(x)dx = \lambda \int dt f(t)e^{-G(y)/2} e^{-G(t)/2},
\eeq

\beqa
\int \int \Delta^2(x,y) f(x) f(y) dx dy &=& \lambda \left[ \int dt f(t) e^{-G(t)/2}\right]^2
\\\nonumber &=&  \lambda \left[ \frac{\beta}{2} \int_0^1 du \frac{P(u)}{1-u} \right]^2
\\\nonumber &=& \lambda^3 \frac{\beta^2}{4} P(1)^2
\eeqa
and $\langle k|\Delta^2|k \rangle=+\lambda^3\frac{\beta}{2} P(1)^2 \left[\int_0^1 du \frac{P(u)^2}{1-u}\right]^{-1}$, which ends up the proof.

\subsection{Limit of $g(k,w)$ when $k\rightarrow +\infty$}
\label{limitg}

We start from an integral representation of $g(k,w)$ defined in (\ref{defg}):
\beqa
\nonumber g(k,w) &=& e^{k w \ln k} \sum_{p=0}^{+\infty} \frac{(-1)^p e^{p w \ln k}}{p! (p+2 k-1)!} \frac{1}{p+k}
\\ \label{g} &=& \frac{i}{2 \pi}  e^{k w \ln k} \int_C dz \int_0^{+\infty} dx e^{S(k,z,x)},
\eeqa
where
\beq
S(k,z,x)=-z-k(x+2\ln(-z))+\frac{k^w}{z}e^{-x},
\eeq
because $(p+k)^{-1}=\int_0^{+\infty} dx \, e^{-(p+k)x}$ and $(p+2k-1)!^{-1}=i/(2\pi) \int_C dz\, e^{-z} (-z)^{-(p+2k)}$. $C$ is a contour in the complex plane such as illustrated in fig. 3.

\begin{figure}
\label{contour}
\begin{center}
\includegraphics[width=6cm]{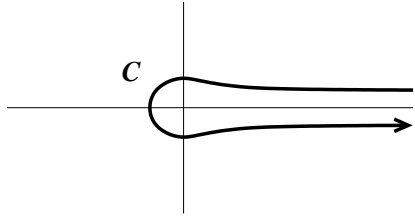}
\caption{The contour $C$ in the complex plane}
\end{center}
\end{figure}

The stationarity equations read
\beqa
\frac{\partial S}{\partial z} & =&  -1-2\frac{k}{z}-\frac{k^w}{z^2} e^{-x}=0,
\\\nonumber \frac{\partial S}{\partial x} & = & -k-\frac{k^w}{z} e^{-x}=0
\eeqa

so that there is a movable saddle-point at
\beqa
\label{movesp}
\left\{ \begin{array}{ll}
x_{sp} &= (w-2) \ln k
\\ z_{sp} &= -k \end{array} \right. .
\eeqa

\subsubsection{The case $w>2$}

In this case the saddle-point (\ref{movesp}) is inside the range of integration. We compute the Hessian at this point:
\beqa
\frac{\partial^2 S}{\partial t^2} &=& 2 \frac{k}{z^2} +2 \frac{k^w}{z^3} e^{-x}=0,
\\\nonumber \frac{\partial^2 S}{\partial x^2} &=& \frac{k^w}{z} e^{-x}=-k,
\\\nonumber \frac{\partial^2 S}{\partial x \partial z} &=& \frac{k^w}{z^2} e^{-x}=1.
\eeqa

So, when $k \rightarrow +\infty$,
\beqa
\int_0^{+\infty} dx e^{S[k,z,x]} &\sim& \int_0^{+\infty} dx \exp \left[S_{sp}-\frac{k}{2}[x-(w-2)\ln k]^2+[x-(w-2)\ln k](z+k) \right]
\\\nonumber &\sim& \sqrt{\frac{2\pi}{k}} e^{S_{sp}} \exp \left[\frac{1}{2k}(z+k)^2 \right].
\eeqa

Then we perform the integration with respect to $z$ with $z+k=-i\epsilon$:

\beqa
\int dz e^{\frac{1}{2k}(z+k)^2} &=& -i \int_{-\infty}^{+\infty} d\epsilon e^{-\frac{\epsilon^2}{2k}}
\\\nonumber &=& -i \sqrt{2\pi k}.
\eeqa

Given that $S_{sp}=-k w \ln k$, we eventually get

\beq
\int \int dx dz e^{S(k,z,x)} \sim -2\pi i e^{-k w \ln k}.
\eeq

It follows that

\beq
g(k,w) \rightarrow_{k\rightarrow +\infty} 1.
\eeq

\subsubsection{The case $w<2$}

In this case the saddle-point (\ref{movesp}) is outside the range of integration. The integral is dominated by $0 \leq x \ll 1$, where

\beq
S(k,z,x)= -z -2 k \ln(-z) +\frac{k^w}{z}-(k+\frac{k^w}{z}) x +O(x^2),
\eeq
so that
\beq
\int_0^{+\infty} dx e^{S(k,z,x)} \sim \frac{\exp \left[-z-2 k \ln(-z)+\frac{k^w}{z}\right]}{k+\frac{k^w}{z}}.
\eeq

We just have to look for the saddle-point of $\Sigma(k,z)=-z-2k\ln(-z)+k^w/z$. The stationarity of $\Sigma$ with respect to $z$ reads

\beq
-1-\frac{2k}{z}-\frac{k^w}{z^2}=0.
\eeq

There are two candidates as a saddle-point:

\beq
z^{\pm}_{sp}=-k \pm\sqrt{k^2-k^w}.
\eeq

It is easy to see that, when $k \rightarrow +\infty$, $\Sigma''(z_{sp}^+)<0$ whereas $\Sigma''(z_{sp}^-)>0$. So on a contour of the shape of \ref{contour} the right saddle-point is $z_{sp}^-$. We have $\Sigma(k,z_{sp}^-)=-2k\ln k+O(k)$, so that 

\beq
\int_C dz \int_0^{+\infty} dx e^{S(k,z,x)}=O\left[e^{-2k\ln k+O(k)}\right].
\eeq

Remembering (\ref{g}), it follows that

\beq
g(k,w)=O\left[ e^{k(w-2)\ln k+O(k)} \right].
\eeq

So $g(k,w)$ goes to 0 when $k \rightarrow +\infty$.

\section{Bipartite case}

\subsection{Computation of the averaged replicated partition function}
\label{replication}

We have two sets of $N$ points each. We introduce the occupation numbers $n_{ij}=0$ or 1, which are constrained by

\beq
\label{condition}
\forall i \in {1, \ldots N}, \,\,\,\,\,\, \sum_{j=1}^{N} n_{ij}=\sum_{j=1}^{N} n_{ji}=1.
\eeq

The length of the matching associated to a choice of the $n_{ij}$ is $L(\{n_{ij}\})=\sum_{i,j} n_{ij} l_{ij}$. The partition function (\ref{parti2}) reads
\beqa
Z &=& \sum_{n_{ij}=0,1} \int_0^{2\pi} \frac{d\lambda_1}{2\pi} \cdots \int_0^{2\pi} \frac{d\lambda_{N}}{2\pi} \int_0^{2\pi} \frac{d\mu_1}{2\pi} \cdots \int_0^{2\pi} \frac{d\mu_{N}}{2\pi}\prod_i \exp \left(i \lambda_i \left(1-\sum_{j=1}^{N} n_{ij} \right) \right)
\\\nonumber && \times \prod_i \exp \left(i \mu_i \left(1-\sum_{j=1}^{N} n_{ji} \right)\right) \exp \left(-\frac{N}{2} \beta \sum_{i,j} n_{ij} l_{ij} \right),
\eeqa
where we enforced (\ref{condition}) using an integral representation of the Kronecker symbol $\delta(p)=\int_0^{2\pi} d\lambda/(2\pi) e^{ip\lambda}$. It follows that
\beqa
Z^n =\int d[\lambda] d[\mu] \exp \left(i \sum_a \sum_i \lambda_i^a+\mu_i^a\right) \prod_{i,j} \left( 1+ \sum_\alpha \exp \left(-\frac{N}{2} p(\alpha) \beta l_{ij} -i \sum_{a\in\alpha} (\lambda_i^{a} + \mu_j^{a})\right) \right)
\eeqa
where $d[\lambda]$ is a shorthand notation for $\prod_{a=1}^n \prod_{i=1}^N d\lambda_i^a/(2\pi)$ (see the beginning of section \ref{nonbipartite} for any precision on the other notations).

As we are interested in the subleading contribution to the free energy, it is important, when averaging over the disorder, to specify the distribution of the lengths we are considering:

\beq
\overline{e^{-N p \beta l_{ij}/2}}=\frac{2}{N}g_p-4\frac{\mu}{N^2}g_p^2+o(\frac{1}{N^2})
\eeq
where $\mu=1$ in the case of the exponential distribution, 0 in the case of the flat distribution. Thus

\beq
\label{eqZ}
\overline{Z^n} = \int d[\lambda] d[\mu] \exp \left(i \sum_a \sum_i \lambda_i^a+\mu_i^a\right) \prod_{i,j} \left( 1+ 2\frac{T_{ij}}{N}-4\mu \frac{R_{ij}}{N^2}+o(\frac{1}{N^2}) \right),
\eeq
where
\beqa
T_{ij}&=& \sum_\alpha g_\alpha \exp \left(-i\sum_{a\in \alpha} \lambda_i^a+\mu_j^a \right),
\\\nonumber R_{ij}&=&\sum_\alpha g_\alpha^2 \exp \left( -i\sum_{a\in \alpha} \lambda_i^a+\mu_j^a \right).
\eeqa

Now we write
\beqa
\prod_{i,j} (1+2\frac{1}{N} T_{ij} -4\frac{\mu}{N^2} R_{ij})&=& \exp\left( \frac{2}{N} \sum_{i,j} T_{ij}\right) \exp \left(-\frac{2}{N^2} \sum_{i,j} T_{ij}^2 \right)
\\\nonumber & & \times \exp \left(-4 \frac{\mu}{N^2} \sum_{i,j} R_{ij} \right) \ldots
\eeqa

Plugged into (\ref{eqZ}), this gives
\beqa
\label{bar}
\lefteqn{\overline{Z^n} = \int d[\lambda] d[\mu] \exp \left(i \sum_a \sum_i \lambda_i^a+\mu_i^a\right) \exp \left(\frac{1}{N} \sum_\alpha g_\alpha (x_\alpha^2+y_\alpha^2) \right)}
\\\nonumber & & \times \exp \left(-\frac{1}{2N^2} \sum_{\alpha,\gamma}' g_{\alpha} g_{\gamma} (x_{\alpha\cup\gamma}^2+ y_{\alpha\cup\gamma}^2) \right) \exp \left(-\frac{\mu}{N^2}\sum_\alpha g_\alpha^2 (x_\alpha^2+y_\alpha^2) \right),
\eeqa
where we set
\beqa
\sum_i \exp \left(-i\sum_{a\in\alpha} \lambda_i^a \right)=(x_\alpha+iy_\alpha)/2,
\\\nonumber \sum_i \exp \left(-i\sum_{a\in\alpha} \mu_i^a\right)=(x_\alpha-iy_\alpha)/2.
\eeqa
In (\ref{bar}), $\sum_{i,j} T_{ij}^2$ gives the only contribution $\sum_{\alpha,\gamma}'$ because the other terms vanish when integrated (remember that for $p$ integer, $\int d\lambda e^{ip\lambda}=0$ unless $p=0$).

Using well known properties of Gaussian integrals, we finally get
\beqa
\lefteqn{\overline{Z^n} = \int d[\lambda] d[\mu] \exp \left(i \sum_a \sum_i \lambda_i^a+\mu_i^a\right) \int \prod_\alpha dX_\alpha dY_\alpha \frac{N}{2\pi g_\alpha} \exp \left(-\frac{N}{2} \sum_\alpha \frac{X_\alpha^2+Y_\alpha^2}{g_\alpha} \right)}
\\\nonumber & &\times \exp \left(\sum_\alpha X_\alpha x_\alpha+Y_\alpha y_\alpha \right) \exp \left(-\frac{1}{2} \sum'_{\alpha, \gamma} \frac{g_\alpha g_\gamma}{g_{\alpha \cup \gamma}^2} (X_{\alpha \cup \gamma}^2+Y_{\alpha \cup \gamma}^2) \right) \exp \left(-\mu\sum_\alpha X_\alpha^2+Y_\alpha^2 \right).
\eeqa

Eventually, expressing $x_\alpha$ and $y_\alpha$ as functions of the $\lambda_i^a$ and $\mu_i^a$ one gets
\beqa
\overline{Z^n} &=& \int \prod_\alpha dX_\alpha dY_\alpha \frac{N}{2\pi g_\alpha} \exp\left(-\frac{N}{2} \sum_\alpha \frac{X_\alpha^2+Y_\alpha^2}{g_\alpha} \right) z[X_\alpha-iY_\alpha]^N z[X_\alpha+iY_\alpha]^N
\\\nonumber & &\times \exp \left(-\frac{1}{2} \sum'_{\alpha, \gamma} \frac{g_\alpha g_\gamma}{g_{\alpha \cup \gamma}^2} (X_{\alpha \cup \gamma}^2+Y_{\alpha \cup \gamma}^2) \right) \exp\left(-\mu \sum_\alpha X_\alpha^2+Y_\alpha^2\right).
\eeqa

\subsection{Computation of $\Delta F_2^1$ (\ref{DF12})}
\label{diff}

Let us recall a result of \cite{mezpar2}

\beq
\label{QG}
Q_p^{sp}=\frac{2}{\beta p} \int_{-\infty}^{+\infty} dl \frac{e^{lp}}{(p-1)!} e^{-G(l)}.
\eeq

Using (\ref{relation}) we see that, when $n \rightarrow 0$ and $\beta\rightarrow +\infty$,
\beqa
\frac{1}{nN\beta} \sum_\alpha (Q_\alpha^{sp})^2 &=& \frac{1}{N\beta} \sum_{p=1}^{+\infty} \frac{(-1)^{p-1}}{p} (Q_p^{sp})^2
\\\nonumber &=& \frac{2}{N\beta^2} \int_{-\infty}^{+\infty} dl \sum_p Q_p^{sp} \frac{(-1)^{p-1}}{p p!} e^{pl} e^{-G(l)}
\\\nonumber &=& \frac{2}{N\beta^2} \int_{-\infty}^{+\infty} dl G(l) \int_l^{+\infty} dt e^{-G(t)} \,\,\,\,\,\,\, \textrm{by (\ref{G})}
\\\nonumber &=& \frac{1}{2N} \int_0^{+\infty} du \ln (1+u) \ln (1+1/u) \,\,\,\,\,\,\, \textrm{by (\ref{asymptG})}
\\\nonumber &=& \frac{1}{N} \zeta(3).
\eeqa

\subsection{Computation of the volume of the hypersurface of saddle-points (\ref{hyp2})}
\label{volume}

The volume is $\int_0^{2\pi} d\theta_1 \ldots \int_0^{2\pi} d\theta_n\, \sqrt{\det g}$ where

\beq
g_{a b}=\sum_{\alpha} \left[ \frac{\partial U_\alpha^{sp}}{\partial \theta_a} \frac{\partial U_\alpha^{sp}}{\partial \theta_b}+\frac{\partial V_\alpha^{sp}}{\partial \theta_a}\frac{\partial V_\alpha^{sp}}{\partial \theta_b} \right].
\eeq

$g$ has a very simple structure: all diagonal elements are equal to $g_0$, all non diagonal elements to $g_1$, with
\beqa
\label{exp1}
g_0 &=&\sum_{p=1}^{+\infty} (U_p^{sp})^2 C_{n-1}^{p-1} \sim_{n\rightarrow 0} \sum_{p=1}^{+\infty} (-1)^{p-1} \frac{(Q_p^{sp})^2}{g_p},
\\\label{exp2} g_1 &=&\sum_{p=2}^{+\infty} (U_p^{sp})^2 C_{n-2}^{p-2} \sim_{n\rightarrow 0} \beta \sum_{p=1}^{+\infty} (-1)^p p^2 (Q_p^{sp})^2+g_0.
\eeqa

Using (\ref{relation}) and (\ref{QG}), one gets
\beqa
g_0 &=& \beta \sum_{p=1}^{+\infty} (-1)^{p-1} p Q_p^{sp} \frac{2}{\beta p} \int_{-\infty}^{+\infty} dl \frac{e^{lp}}{(p-1)!} e^{-G(l)}
\\\nonumber &=& 2\int_{-\infty}^{+\infty} dl G'(l) e^{-G(l)}=2
\eeqa
and
\beqa
g_1-g_0 &=& 2 \sum_{p=1}^{+\infty} (-1)^p p Q_p^{sp} \int_{-\infty}^{+\infty} dl \frac{e^{lp}}{(p-1)!} e^{-G(l)}
\\\nonumber &=& -2 \int_{-\infty}^{+\infty} dl G''(l) e^{-G(l)}=-\frac{2}{\beta}.
\eeqa

The computation of $\det g$ gives

\beq
\det=[g_0(n)-g_1(n)]^{n-1} [g_0(n)+(n-1) g_1(n) ],
\eeq
so that
\beq
\sqrt{\det g}= 1+\frac{n}{2} \left[\frac{g_1}{g_0-g_1}+ \ln (g_0-g_1)\right]+o(n)
\eeq
(here again one must be careful when deriving this result that there might {\sl a priori} be some contribution of $dg_0/dn$ or $dg_1/dn$).

Thus the contribution to the free energy is  
\beqa
\Delta F^3 &=& -\frac{1}{2N\beta} \left[ \frac{g_1}{g_0-g_1}+\ln (g_0-g_1) \right]
\\\nonumber &=_{\beta \rightarrow +\infty}& -\frac{1}{2N}.
\eeqa

M\'ezard and Parisi \cite{mezpar3} had expressions (\ref{exp1}) and (\ref{exp2}) with $(X_p^{sp},Y_p^{sp})$ instead of $(U_p^{sp},V_p^{sp})$, which made them find a wrong $\Delta F^3=-\pi^2/(24N)$.

\end{document}